\documentclass{article}
\usepackage{smc2024}
\usepackage[caption=false, font=footnotesize]{subfig}
\usepackage{paralist}
\usepackage[figure,table]{hypcap}
\usepackage{graphicx}
\usepackage{xparse}
\usepackage{xcolor}
\usepackage{amsmath}
\usepackage{soul}

\DeclareMathOperator*{\argmax}{argmax} 

\author[1]{\mbox{\firstname{Andrea}\lastname{Poltronieri}\email{andrea.poltronieri2@unibo.it}\orcid{0000-0003-3848-7574}}}
\author[1]{\mbox{\firstname{Valentina} \lastname{Presutti}\email{valentina.presutti@unibo.it}\orcid{0000-0002-9380-5160}}}
\author[2,3]{\mbox{\firstname{Martín}\lastname{Rocamora}\email{martin.rocamora@upf.edu}\orcid{0000-0003-3183-9717}}}

\affil[1]{\institution{University of Bologna}\city{Bologna}\country{Italy}\affiliationtype{University}}
\affil[2]{\institution{Universitat Pompeu Fabra}\city{Barcelona}\country{Spain}\affiliationtype{University}}
\affil[3]{\institution{Universidad de la República}\city{Montevideo}\country{Uruguay}\affiliationtype{University}}

\completesetup

\title{ChordSync: Conformer-Based Alignment of Chord Annotations to Music Audio}

\begin{document}

\capstartfalse
\maketitle
\capstarttrue


\begin{abstract}
    In the Western music tradition, chords are the main constituent components of harmony, a fundamental dimension of music. Despite its relevance for several Music Information Retrieval (MIR) tasks, chord-annotated audio datasets are limited and need more diversity. 
    One way to improve those resources is to leverage the large number of chord annotations available online, but this requires aligning them with music audio. However, existing audio-to-score alignment techniques, which typically rely on Dynamic Time Warping (DTW), fail to address this challenge, as they require weakly aligned data for precise synchronisation.
    In this paper, we introduce \emph{ChordSync}, a novel conformer-based model designed to seamlessly align chord annotations with audio, eliminating the need for weak alignment. 
    We also provide a pre-trained model and a user-friendly library, enabling users to synchronise chord annotations with audio tracks effortlessly.
    In this way, ChordSync creates opportunities for harnessing crowd-sourced chord data for MIR, especially in audio chord estimation, thereby facilitating the generation of novel datasets.
    Additionally, our system extends its utility to music education, enhancing music learning experiences by providing accurately aligned annotations, thus enabling learners to engage in synchronised musical practices.
\end{abstract}


\section{Introduction}

Harmony is central to Western music traditions' theoretical and practical foundations. It entails the combination of individual pitches to create chords and their concatenation into sequences to create chord progressions. Therefore, chords, i.e., the simultaneous sounding of two or more pitches, are the primary constituents of harmony, while chord progressions play a vital role in shaping and defining the overall structure of a musical piece. 

Not surprisingly, automatic chord recognition (ACR) from audio, the task of generating a sequence of time-synchronised chord labels given raw audio as input, has been an active research topic in Music Information Retrieval (MIR) for more than two decades~\cite{pauwels2019twentyyears}, with applications including music similarity assessment \cite{dehaas2013similarity, deberardinis2023harmory}, classification \cite{huang2014classification}, and segmentation \cite{pauwels2013segmentation}.

\begin{figure}[t]
    \centering
    \includegraphics[width=\columnwidth]{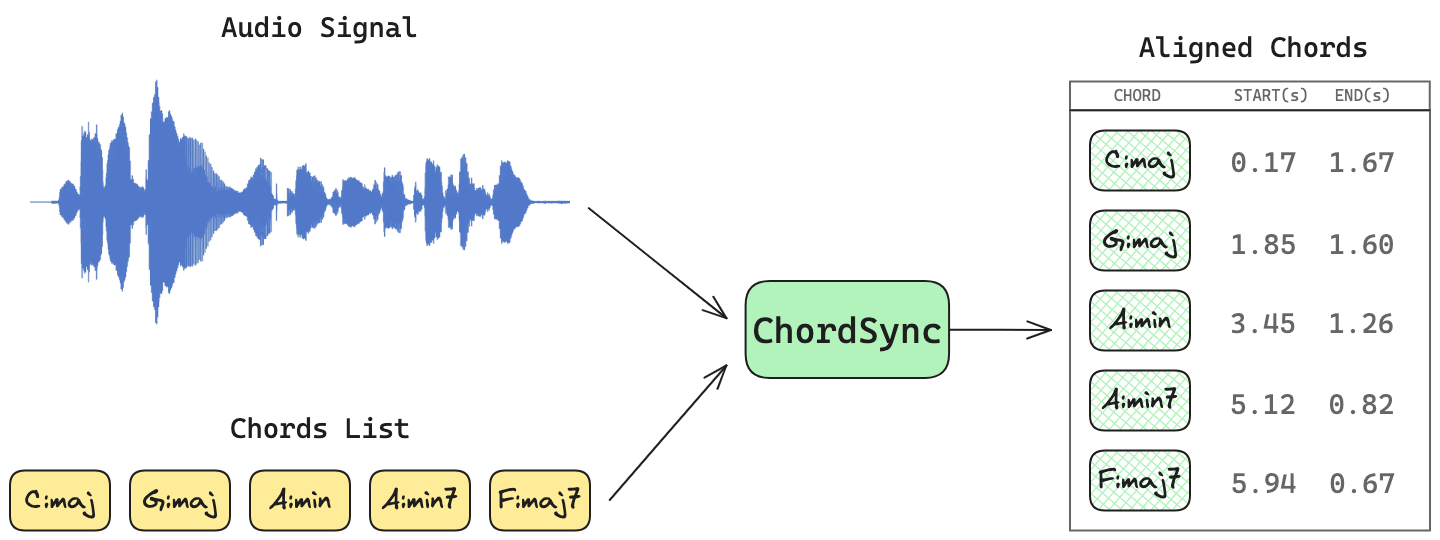}
    \caption{Basic schema of ChordSync: The model processes a list of chords alongside the audio signal, producing time-aligned chords as output.}
    \label{fig:basic-schema}
\end{figure}

The development of ACR systems requires large datasets of audio-aligned chord annotations for training and evaluation. However, the diversity of existing chord annotated datasets is limited. They predominantly feature pop music and exclude a wide array of genres and styles~\cite{pauwels2019twentyyears}.
The lack of diversity is critical since the chord vocabulary differs according to musical style and context, making it difficult to generalise from a limited music sample. Besides, the subjectivity inherent in chord annotations further complicates the ACR task. Musical chords can be annotated at varying levels of granularity and complexity, accounting for global harmony or specific instrument contributions. Additionally, the distinction between harmony and melodic lines is frequently challenging, while interpretations of elements such as arpeggios often lead to divergent annotations.
In \cite{koops2019subjectivity}, authors demonstrate that inter-annotator agreement on the root note in a dataset annotated by four different annotators stands at only $76\%$. Datasets annotated from other perspectives are even rarer, currently comprising only a few dozen tracks.

In recent years, meta-corpora of chord annotations have emerged, such as Chord Corpus (ChoCo) \cite{deberardinis2023choco} and When in Rome (WiR) \cite{gotham2023rome}, which aim to aggregate and standardise different datasets originally available in various formats and annotation styles. In this way, they facilitate the utilisation of large-scale data, which improves diversity and is crucial for training deep-learning models.
However, the availability of audio-aligned annotations within these corpora remains limited. Notably, less than $12\%$ of the $20,000$ annotated tracks in ChoCo are audio-aligned.

\begin{figure*}[t!]
    \centering
    \includegraphics[width=.95\textwidth]{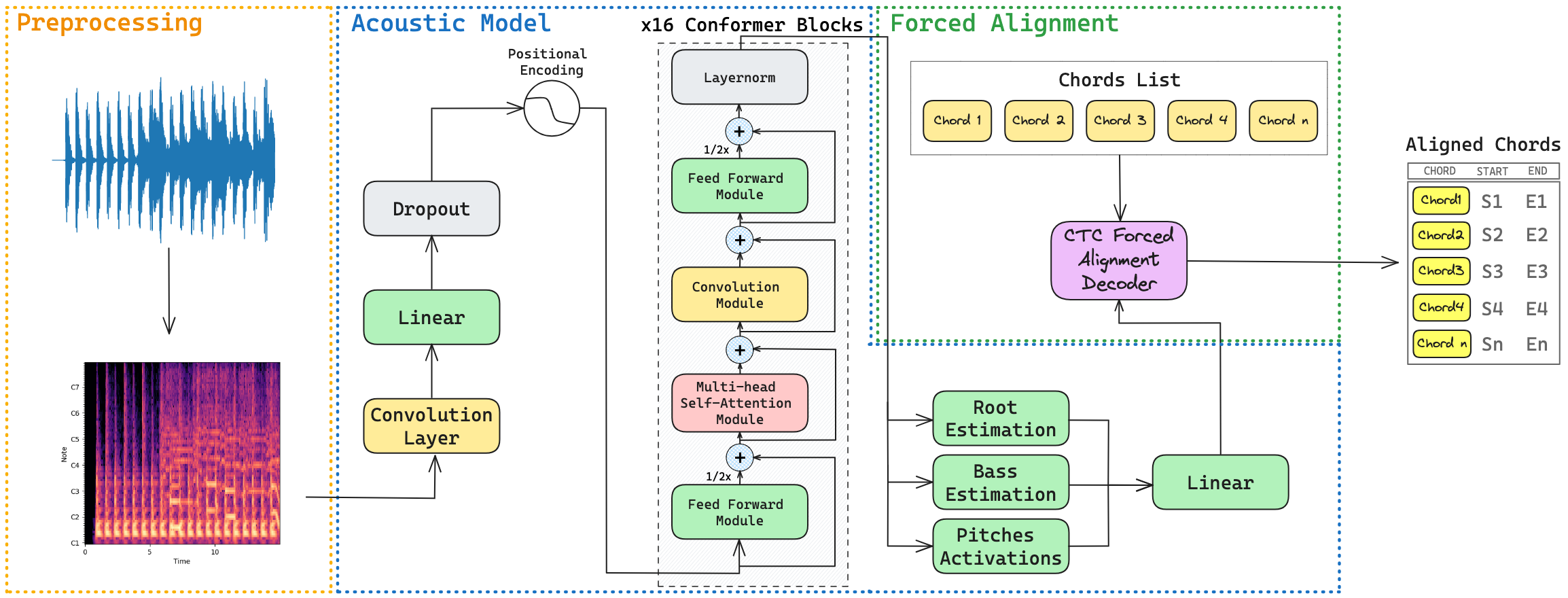}
    \caption{Architecture of \emph{ChordSync}: (i) The audio signal undergoes preprocessing to Constant-Q Transform (\textcolor{orange}{yellow} box); (ii) The preprocessed audio serves as input for training the conformer-based acoustic model (\textcolor{blue}{blue} box); and (iii) The model output probabilities, along with the list of chord labels for alignment, is fed into a CTC forced alignment module (\textcolor{green}{green} box), which outputs the aligned chord labels.}
    \label{fig:model}
\end{figure*}

On the other hand, the internet hosts vast repositories of crowd-sourced chord annotations on platforms such as Ultimate Guitar\footnote{Ultimate Guitar: \url{https://www.ultimate-guitar.com/}}, e-chords\footnote{e-chords: \url{https://www.e-chords.com/}}, and Chordie\footnote{Chordie: \url{https://www.chordie.com/}}, collectively housing millions of annotated songs. 
This multitude offers a great variety in terms of genre distribution, including genres not present in any MIR datasets, such as electronic, metal, hip hop, reggae, and country. 
Moreover, these repositories of harmonic annotations often contain multiple versions of the same song. 
This abundance of versions may offer new avenues for analysis, accommodating the subjectivity and complexity inherent in the annotations, as proposed in \cite{koops2017personalization, koops2019subjectivity}. 
Unfortunately, these annotations lack any timing and duration information, providing solely lists of chords and occasionally lyrics, hindering their reuse for MIR-related tasks.

These challenges underscore the need for systems capable of aligning chord annotations with audio recordings. Yet, to the best of our knowledge, no model has been explicitly developed for this purpose. Existing audio-to-score alignment techniques often rely on Dynamic Time Warping (DTW) algorithms \cite{muller2021synctoolbox},  typically requiring preliminary weak alignment. Such alignment methods are not always feasible for aligning chord annotations to audio, particularly in cases of crowd-sourced data where temporal information is completely lacking.

\subsection{Our Contribution}

In this paper, we address this gap by introducing \emph{ChordSync}, a novel approach that seamlessly aligns chord annotations to audio without requiring any preliminary weak alignment (see Figure~\ref{fig:basic-schema}). Leveraging the power of conformer architecture \cite{gulati2020conformer}, our method paves the way for creating diverse and comprehensive audio-aligned chord annotated datasets based on existing resources. We also provide a pre-trained model and a user-friendly library, enabling users to synchronise chord annotations with audio tracks effortlessly. 
Finally, we showcase the effectiveness of our approach by aligning a sample of tracks taken from Ultimate Guitar.
This can, in turn, benefit other MIR applications, such as music structure analysis, and foster enriched music learning experiences.

The rest of the paper is structured into four main sections: Section \ref{sec:sota} reviews the current state-of-the-art, Section \ref{sec:method} describes the methodology of \emph{ChordSync}, Section \ref{sec:experiments} presents experimental results, and Section \ref{sec:conclusion} offers conclusions and suggests future research directions.

\section{Related Work}
\label{sec:sota}

\subsection{Audio-to-Score Alignment}

The task of aligning audio to symbolic music, commonly known as \emph{audio-to-score alignment} (A2SA), has been primarily addressed by \emph{Dynamic Time Warping (DTW)} algorithms~\cite{morsi2022alignment}, as they are particularly effective for sequence alignment tasks.
Thus, various DTW-based alignment methods have been proposed to align audio with different symbolic music formats, such as MIDI \cite{raffel2016dtw}, often integrating additional techniques and diverse signal representations to improve alignment accuracy~\cite{carabias-orti2015dtw, serrano2016dtw}.

A differentiable variant of DTW, \emph{SoftDTW}, has been recently used as the loss function within neural network architectures, mainly for multi-pitch estimation tasks~\cite{krause2023soft, zeitler2023softdtwimporved}.
However, a general limitation of the DTW-based approaches is their reliance on weak-aligned data to perform the alignment. This requirement renders them unsuitable for contexts without prior alignment information.

Other deep-learning methods have been investigated for audio-to-score alignment, including leveraging automatic transcription techniques~\cite{simonetta2021transcription} and training audio features tailored explicitly for alignment tasks \cite{joder2013features}.

The only previously proposed approach for aligning audio with chord annotations uses Hidden Markov Models (HMM) and is part of an ACR workflow \cite{wu2019nonalignedace}.
Also related to our work is the \emph{Harmonic Change Detector (HCD)}, introduced in \cite{harte2006hcd} and subsequently revisited and improved in \cite{degani2015hcd, ramoneda2020hcdf}, for detecting harmonic changes within the audio signal, including chord changes.
However, the number of harmonic changes within the audio signal often exceeds the number of chord changes, posing challenges for using these algorithms directly for audio-to-chord alignment.

\subsection{Lyrics-to-Audio Alignment}

Another form of alignment pertinent to our work is the audio-to-lyrics alignment task, which seeks to determine the corresponding locations in a song recording of its lyrics at various levels such as line, word, or phoneme~\cite{sharma2019lyricsaudio}.
Existing methods for this task are commonly adapted from automatic speech recognition (ASR)~\cite{gupta2018semisupervised, stoller2019ctclyrics}, despite the inherent complexity of singing voices compared to speech \cite{huang2022improving}, and typically make use of acoustic models trained to recognise the phonetic content of the audio signal at various levels of granularity. 
Some recent works have adopted the Connectionist Temporal Classification (CTC) loss \cite{graves2006ctc}, training the acoustic model in an end-to-end fashion \cite{stoller2019ctclyrics}.

\subsection{Conformer-based Approaches}
The conformer architecture~\cite{gulati2020conformer} has recently emerged in ASR as a novel architecture to effectively model global and local audio dependencies by leveraging a combination of Convolutional Neural Networks (CNNs) and Transformer architectures.
It has showcased remarkable success across various tasks not only in speech \cite{chiu2022selfsupervised} but also in music \cite{won2023foundation}, including melodic transcription \cite{tamer2023violin}, representation learning \cite{duong2022conformerrepresentation}, and music audio enhancement \cite{chae2023conformerenhancement}.

\section{Method}
\label{sec:method}

This section describes \emph{ChordSync}, our proposed conformer-based model for audio-to-chord alignment. It implements an acoustic model for estimating the frame-wise probabilities of chord labels, which are then fed to a forced-alignment decoder, along with the list of chord labels to align. 
Figure \ref{fig:model} illustrates the three primary steps implemented by the model: pre-processing and data augmentation (Section \ref{subsec:preprocessing}), the acoustic model used during training (Section \ref{subsec:model}), and the forced alignment decoder (Section \ref{subsec:alignment}).
The software implementation and a pre-trained model are available on a 
GitHub repository.\footnote{\url{https://github.com/andreamust/ChordSync}}

\subsection{Problem Statement}

Let $X = \{x_1, ..., x_N\}$ be a frame-level sequence of acoustic features extracted from the input audio, where $x_n \in \mathbb{R}^D$ represents a D-dimensional feature vector, and $N$ indicates the total number of frames within the sequence.
Let $C = \{c_1, ..., c_M\}$ be the input list of chord labels encoded into integer values, where $c_m \in \mathbb{Z}^K$, $K$ denotes the size of the chord vocabulary, and $M$ is the length of the chord sequence. 
The list of chord labels is upsampled to match the length of the audio sequence $N$. This upsampling is performed uniformly, assuming each chord has a duration approximately equal to $N/M$. Specifically, each chord label $c_m$ is repeated for approximately $N/M$ frames to produce the sequence $Z = \{z_1, ..., z_N\}$, where $z_m \in \mathbb{Z}^K$. 
Thus, we train an acoustic model to optimise the following equation: 

\begin{equation}
     Z^* = \argmax_z p(Z|X),
    \label{eq:ACE}
\end{equation}
where $Z^*$ represents the optimal sequence of chord labels that maximises the posterior probability $p(Z|X)$, given the input sequence $X$.
Note that $X$ and $Z$ are aligned at the frame level, and $p(X|Z)$ is evaluated by estimating the frame-wise posterior probability $p(x_n|z_n)$.

The output probabilities $p(X|Z)$ from the acoustic model are then fed to a CTC forced alignment decoder, which estimates the best alignment between the sequence of acoustic features $X$ and the list of chord labels $C$:

\begin{equation}
A^* = \argmax_a p(A|X, C),
\label{eq:forced_alignment}
\end{equation}
where $A^*$ represents the optimal alignment between $X$ and $C$ that maximises the posterior probability $p(A|X, C)$.

In this way, the decoder generates the aligned chord labels with respect to the audio signal.  


\subsection{Preprocessing}
\label{subsec:preprocessing}

For the input audio data, a standard pre-processing pipeline is implemented. 
The audio is first resampled to a sampling rate of $22050$ Hz, and a hop size of $2048$ is applied.
Then, the Constant-Q Transform (CQT) features are calculated on $6$ octaves starting from $C1$, with $24$ bins per octave, resulting in a total of $144$ bins.

The audio data used for training undergoes data augmentation by applying (i) time masking and (ii) frequency masking directly to the audio features, as proposed in \emph{SpecAugment} for end-to-end ASR~\cite{park2019augmentation}.
%


During training, each audio excerpt in the training set undergoes augmentation, where either one of the transformations (frequency masking or time masking) or both are applied, and the choice of augmentation technique is determined randomly with equal probability.

\begin{figure}[t!]
    \centering
    \includegraphics[width=1\linewidth]{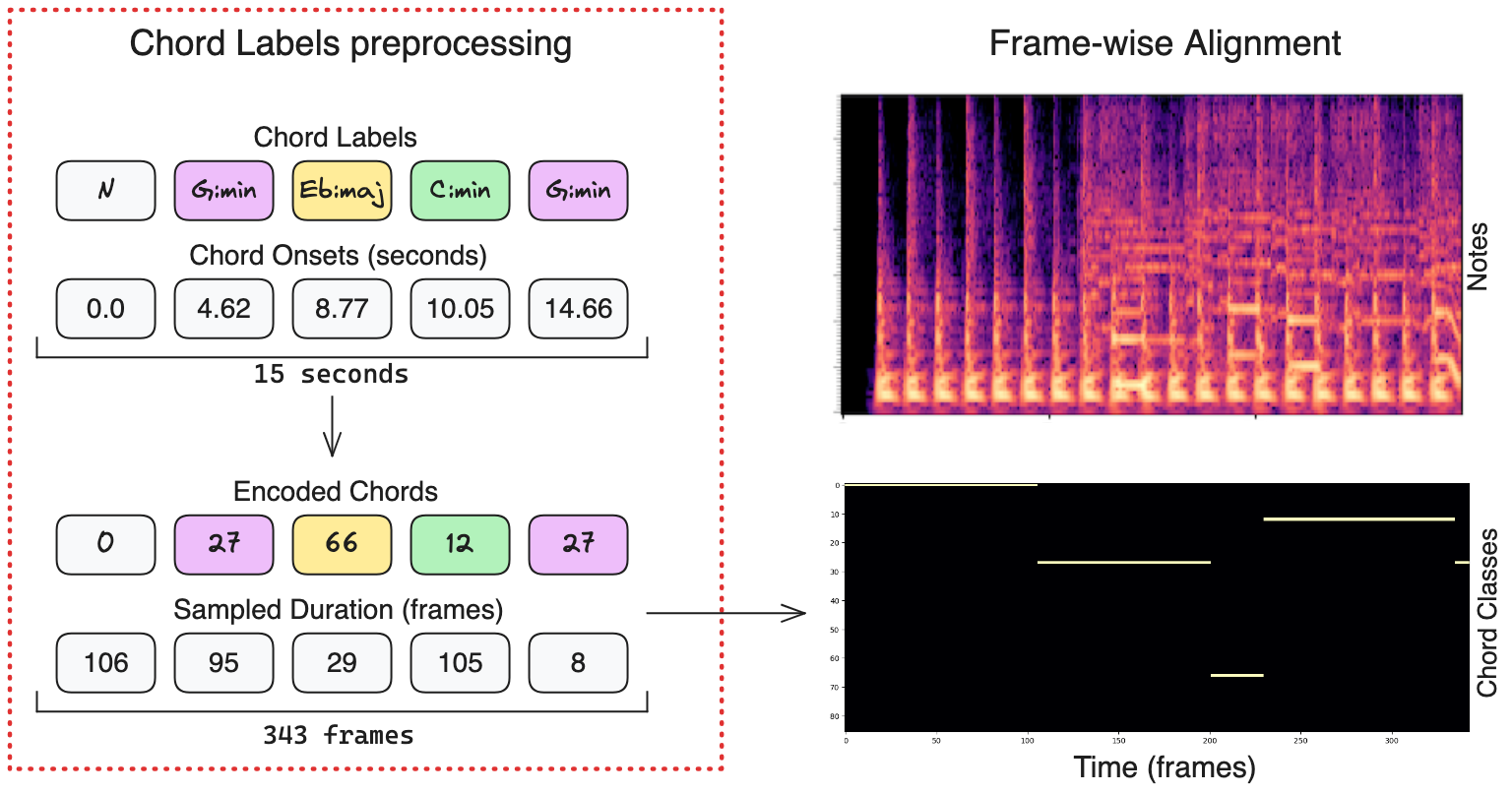}
    \caption{Workflow of the pre-processing applied to the chord labels. Chord labels are numerically encoded and upsampled to match the length of the CQT.}
    \label{fig:chord-preprocessing}
\end{figure}

Chord labels are numerically encoded into integer values and upsampled to match the length of the audio sequence $N$. The upsampling is performed using the \texttt{pumpp} library\footnote{\url{https://github.com/bmcfee/pumpp}.}. 
Figure \ref{fig:chord-preprocessing} shows how chord labels are converted and sampled.
The size of the chord vocabulary $K$ results from the linear combination of the 12 pitches, representing the chromatic scale, with chord qualities such as \texttt{\{maj, min, 7, dim, dim7, hdim7, aug, min7, maj7, maj6, min6, minmaj7, sus2, sus4\}}, plus an additional chord symbol {N} representing silence or no chord.

\subsection{Conformer-based Acoustic Model}
\label{subsec:model}

The acoustic model we adopt is an adaptation of the original Conformer architecture  \cite{gulati2020conformer}, where the audio encoder processes the input through a convolutional module followed by a series of Conformer blocks.

The convolutional module comprises a convolution layer, a fully connected layer, and a dropout layer. The convolutional module serves as the initial feature extractor, capturing local patterns within the input CQT. Dropout regularisation is applied by randomly deactivating units during training to reduce overfitting.
Additionally, we incorporate positional encoding, as proposed in the original transformer architecture paper \cite{vaswani2023attention}.

A Conformer block is composed of four modules stacked together: a feed-forward module, a self-attention module, a convolution module, and a second feed-forward module at the end. 
In the original Conformer paper, the authors explore three different sizes of the Conformer architecture: $S$ (small), $M$ (medium), and $L$ (large), with different numbers of layers, hidden units, and other parameters.
For our implementation, we opt for the $M$ architecture, which comprises a $16$ encoder layer with a dimension of $256$, $4$ attention heads, and a convolutional kernel size of $32$.
While the original paper observed significant improvements when transitioning from the $S$ to the $M$ variant, our experimentation yielded little improvements from $M$ to $L$.

To handle the large dimensionality of the vocabulary, we use an architecture similar to that proposed by \cite{mcfee2017cnn}, in which root notes, bass notes, and all pitch activations of the chord are predicted. 
Subsequently, these probabilities are passed to a feed-forward layer, which converts these three probabilities into the likelihood of the chord with respect to the vocabulary $K$, similarly to what was proposed by \cite{rowe2021acecurriculum}.

For training, we employ cross-entropy loss and optimise using the AdamW optimiser. We utilise a cosine annealed warm restart learning scheduler to manage learning rates. Learning rate schedulers proved effective in training audio data, especially with augmented data \cite{park2019augmentation}. Finally, we applied early stopping by halting the training if the loss failed to decrease for over 20 epochs to prevent overfitting.

\subsection{Forced Alignment}
\label{subsec:alignment}

To estimate the best alignment between the acoustic features $X$ and the chord labels $C$, we utilise the Connectionist Temporal Classification (CTC) objective function \cite{graves2006ctc}, which computes the probability of a given alignment between the input features and output labels. The CTC objective function is defined as follows:
\begin{equation}
p(C|X) = \sum_{A \in \mathcal{A}_{X, C}} p(a_t|X),
\label{eq:CTC}
\end{equation}
where $\mathcal{A}_{X, C}$ denotes the set of all possible alignments that produce the label sequence $C$, and $p(a_t|X)$ represents the probability of alignment $a_t$ given the input sequence $X$.

The probability of alignment $a_t$ given $X$ is computed as the sum of probabilities of all paths $a_t'$ that correspond to $a_t$ after collapsing repeated labels and blank symbols:

\begin{equation}
p(a_t|X) = \prod_{t=1}^{T} p_t(\pi_t|X),
\label{eq:path_probability}
\end{equation}
where $T$ is the length of the alignment, and $p(\pi_t|X)$ is the probability of the $t$-th symbol in the alignment path $\pi$ given the input sequence $X$.

\section{Experiments}
\label{sec:experiments}

\begin{table}[b]
\centering
\resizebox{\columnwidth}{!}{%
\begin{tabular}{l c c c}
\hline
\multicolumn{1}{c}{\textbf{Dataset}} & \textbf{Genre} & \multicolumn{1}{l}{\textbf{\#Tracks}}  & \textbf{Reference} \\ \hline
\textit{Isophonics}           & pop, rock & 300   &    \cite{mauch2009beatles}      \\ 
\textit{Billboard}            & pop       & 740    &   \cite{burgoyne2011expert}      \\ 
\textit{Chordify}             & pop       & 50     &   \cite{chordify2019}      \\ 
\textit{Robbie Williams}      & pop       & 61     &   \cite{di2013automatic}      \\
\textit{Uspop 2002}           & pop       & 195     &  \cite{berenzweig2004large}      \\
\textit{RWC-Pop}              & pop       & 100     &   \cite{goto2002rwc}     \\ 
\textit{Schubert-Winterreise} & classical & 225     &   \cite{weiss2021schubert}     \\  
\textit{Weimar Jazz Database} & jazz      & 456     &   \cite{pfleiderer2017book}     \\ 
\textit{JAAH}                 & jazz      & 113     &   \cite{eremenko2018jaah}     \\  \hline
\textbf{Total}                &          & \textbf{2240}  & \\ \hline
\end{tabular}%
}
\caption{Dataset utilised for all experiments. All datasets are sourced from ChoCo \cite{deberardinis2023choco}.}
\label{tab:dataset-table}
\end{table}

\begin{table}[ht!]
\centering
\resizebox{\columnwidth}{!}{%
\begin{tabular}{ l l c c c }
\hline
\multicolumn{1}{c}{\textbf{Method}} & \textbf{Genre} & \textbf{Precision \(\uparrow\)} & \textbf{Recall \(\uparrow\)} & \textbf{F1 Score \(\uparrow\)} \\ \hline
\textit{HCDF}               & pop/rock  & 0.4999 & 0.6334 & 0.5269 \\ 
\textit{HCDF}               & classical & 0.4454 & 0.6220  & 0.5191 \\ 
\textit{HCDF}               & jazz      & 0.4911 & 0.7749 & 0.5857 \\ 
\textit{HCDF}               & all       & 0.4953 & 0.6508 & 0.5323 \\ \hline
\textit{ChordSync}   & pop/rock & 0.8847 & 0.8335 & 0.8553 \\ 
\textit{ChordSync}   & classical & 0.6008 & 0.5917 & 0.5951 \\ 
\textit{ChordSync}   & jazz & 0.4663 & 0.4129 & 0.4350 \\ 
\textit{ChordSync}   & all       & \textbf{0.8895} & \textbf{0.8420} & \textbf{0.8621} \\ \hline
\end{tabular}%
}
\caption{Precision, Recall, and F1 Score for the \emph{HCDF} method \cite{ramoneda2020hcdf} and the proposed \emph{ChordSync} model.}
\label{tab:hcdf-table}
\end{table}

Due to the lack of established methodologies to address the chord-to-score alignment task, conducting a comparative evaluation with existing state-of-the-art techniques presents some challenges. Therefore, to gauge the effectiveness of the proposed methodologies, we use alternative approaches performing analogous albeit slightly dissimilar methods for comparison and conduct two different experiments. The first aims to evaluate the model's capability in detecting chord boundaries, while the second compares it to a traditional DTW-based alignment. 

All experiments were carried out using a subset of ChoCo \cite{deberardinis2023choco}, which offers a standardised version of chord annotations sourced from various datasets. 
Specifically, only ChoCo partitions annotated on audio, i.e. expressing temporal information such as onsets and duration in seconds, were considered. 
Table \ref{tab:dataset-table} presents a summary of all ChoCo partitions employed for training and evaluation.

Audio files corresponding to each ChoCo annotation were obtained automatically from the available metadata in the original datasets. This was necessary as only a small portion of the datasets offer external links to the original audio sources used for chord annotation. 
Since the automatic retrieval process depends on sometimes sparse and incomplete metadata, the validity of the audio files was manually verified on randomly selected samples.
The complete dataset consists of $2240$ audio tracks, encompassing four distinct music genres: pop, rock, classical, and jazz. However, it is noteworthy to observe a significant imbalance in the dataset, with the pop/rock genre comprising over $65\%$ of the total tracks.

Audio data is segmented into intervals of $15$ seconds duration, with a $3$-second overlap between each segment and the preceding one, yielding a corpus of $31909$ segments. We split these segments into train, validation, and test sets with proportions of $65-20-15$.
Importantly, when a segment from a particular song is included in the train set, we ensure that no segments from the same song are included in either the validation or test sets.

\subsection{Chord Changes Detection Evaluation}

The first comparison is conducted with the Harmonic Change Detection algorithm~\cite{harte2006hcd}, which specialises in detecting harmonic changes on an audio signal. 
These algorithms are typically evaluated by assessing their capacity to detect the onsets of annotated chords within the identified harmonic changes, often employing standard metrics such as Precision, Recall, and F1 Score.

However, by their intrinsic design, HCD algorithms extract every harmonic variation present in the audio signal. \cite{harte2006hcd} and \cite{ramoneda2020hcdf} provide two distinct implementations of this algorithm, each optimising either the F1 score or precision. 
The number of harmonic changes varies significantly depending on the chosen algorithm implementation, but in general, it far exceeds that of chord changes.

In contrast, \emph{ChordSync} extracts the number of chord changes  of the list of chords passed to the CTC decoder.
Table \ref{tab:hcdf-table} presents a comparative analysis between the HCD algorithm in \cite{ramoneda2020hcdf} and \emph{ChordSync}. 
A harmonic change match is defined in a 0.3 seconds window between the predicted and the ground-truth onsets. 

Our method demonstrates notable efficacy in chord change extraction, substantially increasing all the performance measures considered. This performance improvement stems from the model's inherent design, which optimises the alignment between the audio signal and the provided sequence of chords. However, performance decreases in the less represented genres within the dataset, such as jazz and classical.

\subsection{Alignment Evaluation}

\begin{table*}[t]
\centering
\resizebox{\textwidth}{!}{%
\begin{tabular}{llccccc}
\hline
\multicolumn{1}{c}{\textbf{Method}} &
  \multicolumn{1}{c}{\textbf{Dataset}} &
  \textbf{Percentage Correct \(\uparrow\)} &
  \textbf{Median Absolute Error \(\downarrow\)} &
  \textbf{Average Absolute Error \(\downarrow\)} &
  \textbf{Perceptual \(\uparrow\)} \\ \hline
\textit{DTW}                & schubert-winterreise & 0.8621 & 0.0661 & 0.2088 & 0.7895 \\ 
\textit{ChordSync} & schubert-winterreise & 0.8245 & 0.2641 &0.2512  & 0.7230 \\ \hline
\textit{ChordSync} & all                  & 0.8664   & 0.4224 & 0.5001 & 0.7900 \\ \hline
\end{tabular}%
}
\caption{Performance of ChordSync on the Schubert-Winterreise dataset \cite{weiss2021schubert} compared to a standard DTW approach performed using the \emph{SyncToolbox} library (first two rows). Additionally, performance metrics of the ChordSync method applied across all datasets are presented. Metrics are computed with the alignment metrics from the \texttt{mir\_eval} library.}


\label{tab:dtw-evaluation}
\end{table*}

Evaluating audio-to-score or audio-to-lyrics alignment entails comparing predicted and ground truth timestamps to measure their temporal differences \cite{mauch2010alignment, fujihara2011alignment}. 
This comparison typically occurs pairwise and involves calculating metrics such as the median absolute error in seconds and the percentage of overlapping segments. This approach offers a straightforward means of assessing alignment accuracy and determining the effectiveness of alignment methods for practical applications.

Furthermore, perceptually-grounded metrics for evaluating lyrics-to-audio alignment systems have been recently introduced \cite{masclef2021user}. 
These metrics were fine-tuned on data collected through a user Karaoke-like experiment, reflecting human judgement of how ``synchronous'' lyrics and audio stimuli are perceived in that setup.

All the metrics described above are implemented in the \texttt{mir\_eval} library \cite{raffel2014mireval}, providing a standardised and accessible means for conducting evaluations in audio alignment.
Given its similarities with other alignment tasks and the perceptual considerations involved, the same metrics are suitable for evaluating audio-to-chord alignment.

We compare the performance of \emph{ChordSync} and a conventional DTW-based approach using the \emph{SyncToolbox} library \cite{muller2021synctoolbox}, which offers a diverse array of DTW-based implementations.
The evaluation of this type of approach requires both symbolic sequences weakly aligned to audio, which are a prerequisite for the alignment, and ground truth annotations strong aligned to audio for evaluation. 
To our current knowledge, such annotations are exclusively found within the Schubert Winterreise dataset \cite{weiss2021schubert}. 
Consequently, the evaluation of this approach is constrained to a limited number of pieces and to the \emph{classical} genre.

To perform the alignment between audio and chord annotations, the chord annotations were first decomposed into their constituent notes, each of which was then associated with the chord's symbolic onsets. The audio data underwent pre-processing using chroma and DLNCO features, known for their effectiveness in alignment tasks \cite{ewert2009dlnco}.
Finally, alignment was carried out utilising memory-restricted multi-scale DTW (MrMsDTW) \cite{muller2006multiscale, pratzlich2016mr-dtw}.

Table \ref{tab:dtw-evaluation} shows the performance of the proposed model on the Schubert Winterreise dataset compared to a standard DTW approach, along with the broader performance metrics of the ChordSync method applied across all datasets (c.f. Table \ref{tab:dataset-table}).
This evaluation demonstrates that the proposed model accurately detects chord changes and achieves alignment performance comparable to that of a DTW-based approach. 
Conversely, the evaluation conducted solely on a subset of the Winterreise dataset demonstrates performance comparable to DTW, albeit slightly lower.  
However, this data highlights the model's strong generalisation capabilities, as it effectively aligns songs from a genre that was statistically rare in the training data due to its limited size.

Even so, it is worth noting that the proposed model achieves these results without relying on weak-aligned data, which is a requirement for DTW-based approaches.

\section{Discussion and Conclusion}
\label{sec:conclusion}

In this paper, we introduce \emph{ChordSync}, a novel model for audio-to-chord alignment that harnesses the capabilities of the Conformer architecture \cite{gulati2020conformer}. Our proposed method attains performance levels comparable to DTW algorithms in the audio-to-chord alignment task without requiring any pre-existing alignment as in the DTW approaches. 
Therefore, our method facilitates the creation of diverse and comprehensive datasets featuring synchronised audio and chord annotations by exploiting existing resources, such as crowd-sourced online chord annotations, which typically lack timing and duration information. In order to do that, we offer a pre-trained model and a user-friendly library, empowering users to synchronise chord annotations with audio tracks effortlessly.

The primary limitation of the proposed approach stems from its reliance on an acoustic model trained using a simplified vocabulary of chord labels (see Section \ref{sec:method}) because the model's performance is contingent upon the vocabulary size. If a chord is absent from the chord vocabulary, it will inevitably be approximated by the existing label of the most similar chord in the vocabulary. 
However, if the two consecutive chord symbols match, the alignment gets more challenging for the CTC decoder.
Although the results indicate that the decoder can handle such scenarios, using alternative chord encoding might yield better performance. 

Furthermore, investigating alternative chord encoding could make the model key-agnostic, a feature lacking in the current model, which is not specifically designed to handle discrepancies in key between the chord labels and the audio signal.


\begin{acknowledgments}
  This project has received funding from the European Union's Horizon 2020 research and innovation programme under grant agreement No 101004746.
\end{acknowledgments} 
	

\bibliography{bibliography}

\begin{thebibliography}{10}
\providecommand{\url}[1]{#1}
\csname url@samestyle\endcsname
\providecommand{\newblock}{\relax}
\providecommand{\bibinfo}[2]{#2}
\providecommand{\BIBentrySTDinterwordspacing}{\spaceskip=0pt\relax}
\providecommand{\BIBentryALTinterwordstretchfactor}{4}
\providecommand{\BIBentryALTinterwordspacing}{\spaceskip=\fontdimen2\font plus
\BIBentryALTinterwordstretchfactor\fontdimen3\font minus \fontdimen4\font\relax}
\providecommand{\BIBforeignlanguage}[2]{{%
\expandafter\ifx\csname l@#1\endcsname\relax
\typeout{** WARNING: IEEEtran.bst: No hyphenation pattern has been}%
\typeout{** loaded for the language `#1'. Using the pattern for}%
\typeout{** the default language instead.}%
\else
\language=\csname l@#1\endcsname
\fi
#2}}
\providecommand{\BIBdecl}{\relax}
\BIBdecl

\bibitem{pauwels2019twentyyears}
J.~Pauwels, K.~O'Hanlon, E.~G{\'{o}}mez, and M.~B. Sandler, ``20 years of automatic chord recognition from audio,'' in \emph{Proceedings of the 20th International Society for Music Information Retrieval Conference, {ISMIR} 2019, Delft, The Netherlands, November 4-8, 2019}, A.~Flexer, G.~Peeters, J.~Urbano, and A.~Volk, Eds., 2019, pp. 54--63.

\bibitem{dehaas2013similarity}
W.~B. de~Haas, F.~Wiering, and R.~C. Veltkamp, ``A geometrical distance measure for determining the similarity of musical harmony,'' \emph{Int. J. Multim. Inf. Retr.}, vol.~2, no.~3, pp. 189--202, 2013.

\bibitem{deberardinis2023harmory}
J.~de~Berardinis, A.~Mero{\~{n}}o{-}Pe{\~{n}}uela, A.~Poltronieri, and V.~Presutti, ``The harmonic memory: a knowledge graph of harmonic patterns as a trustworthy framework for computational creativity,'' in \emph{Proceedings of the {ACM} Web Conference 2023, {WWW} 2023, Austin, TX, USA, 30 April 2023 - 4 May 2023}, Y.~Ding, J.~Tang, J.~F. Sequeda, L.~Aroyo, C.~Castillo, and G.~Houben, Eds.\hskip 1em plus 0.5em minus 0.4em\relax {ACM}, 2023, pp. 3873--3882.

\bibitem{huang2014classification}
Y.~Huang, S.~Lin, H.~Wu, and Y.~Li, ``Music genre classification based on local feature selection using a self-adaptive harmony search algorithm,'' \emph{Data Knowl. Eng.}, vol.~92, pp. 60--76, 2014.

\bibitem{pauwels2013segmentation}
J.~Pauwels, F.~Kaiser, and G.~Peeters, ``Combining harmony-based and novelty-based approaches for structural segmentation,'' in \emph{International Society for Music Information Retrieval Conference}, 2013.

\bibitem{koops2019subjectivity}
H.~V. Koops, W.~B. De~Haas, J.~A. Burgoyne, J.~Bransen, A.~Kent-Muller, and A.~Volk, ``\BIBforeignlanguage{en}{Annotator subjectivity in harmony annotations of popular music},'' \emph{\BIBforeignlanguage{en}{Journal of New Music Research}}, vol.~48, no.~3, p. 232–252, may 2019.

\bibitem{deberardinis2023choco}
J.~de~Berardinis, A.~Mero{\~{n}}o-Pe{\~{n}}uela, A.~Poltronieri, and V.~Presutti, ``Choco: a chord corpus and a data transformation workflow for musical harmony knowledge graphs,'' \emph{Scientific Data}, vol.~10, no.~1, p. 641, Sep 2023.

\bibitem{gotham2023rome}
M.~Gotham, G.~Micchi, N.~N. L{\'o}pez, and M.~Sailor, ``When in rome: a meta-corpus of functional harmony,'' \emph{Transactions of the International Society for Music Information Retrieval}, vol.~6, no.~1, 2023.

\bibitem{koops2017personalization}
\BIBentryALTinterwordspacing
H.~V. Koops, W.~B. de~Haas, J.~Bransen, and A.~Volk, ``Chord label personalization through deep learning of integrated harmonic interval-based representations,'' \emph{CoRR}, vol. abs/1706.09552, 2017. [Online]. Available: \url{http://arxiv.org/abs/1706.09552}
\BIBentrySTDinterwordspacing

\bibitem{muller2021synctoolbox}
M.~Müller, Y.~Özer, M.~Krause, T.~Prätzlich, and J.~Driedger, ``Sync toolbox: A python package for efficient, robust, and accurate music synchronization,'' \emph{Journal of Open Source Software}, vol.~6, no.~64, p. 3434, 2021.

\bibitem{gulati2020conformer}
A.~Gulati, J.~Qin, C.~Chiu, N.~Parmar, Y.~Zhang, J.~Yu, W.~Han, S.~Wang, Z.~Zhang, Y.~Wu, and R.~Pang, ``Conformer: Convolution-augmented transformer for speech recognition,'' in \emph{Interspeech 2020, 21st Annual Conference of the International Speech Communication Association, Virtual Event, Shanghai, China, 25-29 October 2020}, H.~Meng, B.~Xu, and T.~F. Zheng, Eds.\hskip 1em plus 0.5em minus 0.4em\relax {ISCA}, 2020, pp. 5036--5040.

\bibitem{morsi2022alignment}
A.~Morsi and X.~Serra, ``Bottlenecks and solutions for audio to score alignment research,'' in \emph{Proceedings of the 23rd International Society for Music Information Retrieval Conference, {ISMIR} 2022, Bengaluru, India, December 4-8, 2022}, P.~Rao, H.~A. Murthy, A.~Srinivasamurthy, R.~M. Bittner, R.~C. Repetto, M.~Goto, X.~Serra, and M.~Miron, Eds., 2022, pp. 272--279.

\bibitem{raffel2016dtw}
C.~Raffel and D.~P.~W. Ellis, ``Optimizing dtw-based audio-to-midi alignment and matching,'' in \emph{2016 IEEE International Conference on Acoustics, Speech and Signal Processing (ICASSP)}, 2016, pp. 81--85.

\bibitem{carabias-orti2015dtw}
J.~J. Carabias{-}Orti, F.~J. Rodr{\'{\i}}guez{-}Serrano, P.~Vera{-}Candeas, N.~Ruiz{-}Reyes, and F.~J. Ca{\~{n}}adas{-}Quesada, ``An audio to score alignment framework using spectral factorization and dynamic time warping,'' in \emph{Proceedings of the 16th International Society for Music Information Retrieval Conference, {ISMIR} 2015, M{\'{a}}laga, Spain, October 26-30, 2015}, M.~M{\"{u}}ller and F.~Wiering, Eds., 2015, pp. 742--748.

\bibitem{serrano2016dtw}
F.~J. Rodriguez-Serrano, J.~J. Carabias-Orti, P.~Vera-Candeas, and D.~Martinez-Munoz, ``Tempo driven audio-to-score alignment using spectral decomposition and online dynamic time warping,'' \emph{ACM Trans. Intell. Syst. Technol.}, vol.~8, no.~2, oct 2016.

\bibitem{krause2023soft}
M.~Krause, C.~Wei{\ss}, and M.~M{\"u}ller, ``Soft dynamic time warping for multi-pitch estimation and beyond,'' in \emph{ICASSP 2023-2023 IEEE International Conference on Acoustics, Speech and Signal Processing (ICASSP)}.\hskip 1em plus 0.5em minus 0.4em\relax IEEE, 2023, pp. 1--5.

\bibitem{zeitler2023softdtwimporved}
J.~Zeitler, S.~Deniffel, M.~Krause, and M.~M{\"{u}}ller, ``Stabilizing training with soft dynamic time warping: {A} case study for pitch class estimation with weakly aligned targets,'' in \emph{Proceedings of the 24th International Society for Music Information Retrieval Conference, {ISMIR} 2023, Milan, Italy, November 5-9, 2023}, A.~Sarti, F.~Antonacci, M.~Sandler, P.~Bestagini, S.~Dixon, B.~Liang, G.~Richard, and J.~Pauwels, Eds., 2023, pp. 433--439.

\bibitem{simonetta2021transcription}
F.~Simonetta, S.~Ntalampiras, and F.~Avanzini, ``Audio-to-score alignment using deep automatic music transcription,'' in \emph{2021 IEEE 23rd International Workshop on Multimedia Signal Processing (MMSP)}, 2021, pp. 1--6.

\bibitem{joder2013features}
C.~Joder, S.~Essid, and G.~Richard, ``Learning optimal features for polyphonic audio-to-score alignment,'' \emph{IEEE Transactions on Audio, Speech, and Language Processing}, vol.~21, no.~10, pp. 2118--2128, 2013.

\bibitem{wu2019nonalignedace}
Y.~Wu, T.~Carsault, and K.~Yoshii, ``Automatic chord estimation based on a frame-wise convolutional recurrent neural network with non-aligned annotations,'' in \emph{27th European Signal Processing Conference, {EUSIPCO} 2019, {A} Coru{\~{n}}a, Spain, September 2-6, 2019}.\hskip 1em plus 0.5em minus 0.4em\relax {IEEE}, 2019, pp. 1--5.

\bibitem{harte2006hcd}
C.~Harte, M.~Sandler, and M.~Gasser, ``Detecting harmonic change in musical audio,'' in \emph{Proceedings of the 1st ACM Workshop on Audio and Music Computing Multimedia}, ser. AMCMM '06.\hskip 1em plus 0.5em minus 0.4em\relax New York, NY, USA: Association for Computing Machinery, 2006, p. 21–26.

\bibitem{degani2015hcd}
A.~Degani, M.~Dalai, R.~Leonardi, and P.~Migliorati, ``Harmonic change detection for musical chords segmentation,'' in \emph{2015 {IEEE} International Conference on Multimedia and Expo, {ICME} 2015, Turin, Italy, June 29 - July 3, 2015}.\hskip 1em plus 0.5em minus 0.4em\relax {IEEE} Computer Society, 2015, pp. 1--6.

\bibitem{ramoneda2020hcdf}
P.~Ramoneda~Franco and G.~Bernardes~de Almeida, ``Revisiting harmonic change detection,'' in \emph{Audio Engineering Society Convention}, vol. 149, oct 2020.

\bibitem{sharma2019lyricsaudio}
B.~Sharma, C.~Gupta, H.~Li, and Y.~Wang, ``Automatic lyrics-to-audio alignment on polyphonic music using singing-adapted acoustic models,'' in \emph{ICASSP 2019 - 2019 IEEE International Conference on Acoustics, Speech and Signal Processing (ICASSP)}, 2019, pp. 396--400.

\bibitem{gupta2018semisupervised}
C.~Gupta, R.~Tong, H.~Li, and Y.~Wang, ``Semi-supervised lyrics and solo-singing alignment,'' in \emph{International Society for Music Information Retrieval Conference}, 2018.

\bibitem{stoller2019ctclyrics}
D.~Stoller, S.~Durand, and S.~Ewert, ``End-to-end lyrics alignment for polyphonic music using an audio-to-character recognition model,'' 2019.

\bibitem{huang2022improving}
J.~Huang, E.~Benetos, and S.~Ewert, ``Improving lyrics alignment through joint pitch detection,'' 2022.

\bibitem{graves2006ctc}
A.~Graves, S.~Fern\'{a}ndez, F.~Gomez, and J.~Schmidhuber, ``Connectionist temporal classification: labelling unsegmented sequence data with recurrent neural networks,'' in \emph{Proceedings of the 23rd International Conference on Machine Learning}, ser. ICML '06.\hskip 1em plus 0.5em minus 0.4em\relax New York, NY, USA: Association for Computing Machinery, 2006, p. 369–376.

\bibitem{chiu2022selfsupervised}
C.-C. Chiu, J.~Qin, Y.~Zhang, J.~Yu, and Y.~Wu, ``Self-supervised learning with random-projection quantizer for speech recognition,'' 2022.

\bibitem{won2023foundation}
M.~Won, Y.-N. Hung, and D.~Le, ``A foundation model for music informatics,'' 2023.

\bibitem{tamer2023violin}
N.~C. Tamer, Y.~{\"O}zer, M.~M{\"u}ller, and X.~Serra, ``High-resolution violin transcription using weak labels,'' in \emph{Ismir 2023 Hybrid Conference}, 2023.

\bibitem{duong2022conformerrepresentation}
Q.~T. Duong, D.~H. Nguyen, B.~T. Ta, N.~M. Le, and V.~H. Do, ``Improving self-supervised audio representation based on contrastive learning with conformer encoder,'' in \emph{Proceedings of the 11th International Symposium on Information and Communication Technology}, ser. SoICT '22.\hskip 1em plus 0.5em minus 0.4em\relax New York, NY, USA: Association for Computing Machinery, 2022, p. 270–275.

\bibitem{chae2023conformerenhancement}
Y.~Chae, J.~Koo, S.~Lee, and K.~Lee, ``Exploiting time-frequency conformers for music audio enhancement,'' in \emph{Proceedings of the 31st ACM International Conference on Multimedia}, ser. MM '23.\hskip 1em plus 0.5em minus 0.4em\relax New York, NY, USA: Association for Computing Machinery, 2023, p. 2362–2370.

\bibitem{park2019augmentation}
D.~S. Park, W.~Chan, Y.~Zhang, C.-C. Chiu, B.~Zoph, E.~D. Cubuk, and Q.~V. Le, ``Specaugment: A simple data augmentation method for automatic speech recognition,'' in \emph{Interspeech 2019}.\hskip 1em plus 0.5em minus 0.4em\relax ISCA, Sep. 2019.

\bibitem{vaswani2023attention}
A.~Vaswani, N.~Shazeer, N.~Parmar, J.~Uszkoreit, L.~Jones, A.~N. Gomez, L.~Kaiser, and I.~Polosukhin, ``Attention is all you need,'' 2023.

\bibitem{mcfee2017cnn}
\BIBentryALTinterwordspacing
B.~McFee and J.~P. Bello, ``Structured training for large-vocabulary chord recognition,'' in \emph{Proceedings of the 18th International Society for Music Information Retrieval Conference, {ISMIR} 2017, Suzhou, China, October 23-27, 2017}, S.~J. Cunningham, Z.~Duan, X.~Hu, and D.~Turnbull, Eds., 2017, pp. 188--194. [Online]. Available: \url{https://ismir2017.smcnus.org/wp-content/uploads/2017/10/77\_Paper.pdf}
\BIBentrySTDinterwordspacing

\bibitem{rowe2021acecurriculum}
\BIBentryALTinterwordspacing
L.~O. Rowe and G.~Tzanetakis, ``Curriculum learning for imbalanced classification in large vocabulary automatic chord recognition,'' in \emph{Proceedings of the 22nd International Society for Music Information Retrieval Conference, {ISMIR} 2021, Online, November 7-12, 2021}, J.~H. Lee, A.~Lerch, Z.~Duan, J.~Nam, P.~Rao, P.~van Kranenburg, and A.~Srinivasamurthy, Eds., 2021, pp. 586--593. [Online]. Available: \url{https://archives.ismir.net/ismir2021/paper/000073.pdf}
\BIBentrySTDinterwordspacing

\bibitem{mauch2009beatles}
M.~Mauch, C.~Cannam, M.~Davies, S.~Dixon, C.~Harte, S.~Kolozali, D.~Tidhar, and M.~Sandler, ``{OMRAS2 metadata project 2009},'' in \emph{International Society for Music Information Retrieval (ISMIR)}, 2009.

\bibitem{burgoyne2011expert}
J.~A. Burgoyne, J.~Wild, and I.~Fujinaga, ``{An Expert Ground Truth Set for Audio Chord Recognition and Music Analysis},'' in \emph{International Society for Music Information Retrieval (ISMIR)}, vol.~11, 2011, pp. 633--638.

\bibitem{chordify2019}
H.~V. Koops, B.~de~Haas, J.~A. Burgoyne, J.~Bransen, A.~Kent-Muller, and A.~Volk, ``Annotator subjectivity in harmony annotations of popular music,'' \emph{Journal of New Music Research}, vol.~48, no.~3, pp. 232--252, 2019.

\bibitem{di2013automatic}
B.~Di~Giorgi, M.~Zanoni, A.~Sarti, and S.~Tubaro, ``Automatic chord recognition based on the probabilistic modeling of diatonic modal harmony,'' in \emph{Proceedings of the 8th International Workshop on Multidimensional Systems}.\hskip 1em plus 0.5em minus 0.4em\relax VDE, 2013, pp. 1--6.

\bibitem{berenzweig2004large}
A.~Berenzweig, B.~Logan, D.~P. Ellis, and B.~Whitman, ``A large-scale evaluation of acoustic and subjective music-similarity measures,'' \emph{Computer Music Journal}, pp. 63--76, 2004.

\bibitem{goto2002rwc}
M.~Goto, H.~Hashiguchi, T.~Nishimura, and R.~Oka, ``{RWC Music Database: Popular, Classical and Jazz Music Databases},'' in \emph{International Society for Music Information Retrieval (ISMIR)}, vol.~2, 2002, pp. 287--288.

\bibitem{weiss2021schubert}
C.~Wei{\ss}, F.~Zalkow, V.~Arifi-M{\"u}ller, M.~M{\"u}ller, H.~V. Koops, A.~Volk, and H.~G. Grohganz, ``{Schubert Winterreise dataset: A multimodal scenario for music analysis},'' \emph{Journal on Computing and Cultural Heritage (JOCCH)}, vol.~14, no.~2, pp. 1--18, 2021.

\bibitem{pfleiderer2017book}
M.~Pfleiderer, K.~Frieler, J.~Abe{\ss}er, W.-G. Zaddach, and B.~Burkhart, Eds., \emph{{I}nside the {J}azzomat - {N}ew {P}erspectives for {J}azz {R}esearch}.\hskip 1em plus 0.5em minus 0.4em\relax Schott Campus, 2017.

\bibitem{eremenko2018jaah}
\BIBentryALTinterwordspacing
V.~Eremenko, E.~Demirel, B.~Bozkurt, and X.~Serra, ``{JAAH: Audio-aligned jazz harmony dataset},'' Jun. 2018. [Online]. Available: \url{https://doi.org/10.5281/zenodo.1290737}
\BIBentrySTDinterwordspacing

\bibitem{mauch2010alignment}
M.~Mauch, H.~Fujihara, and M.~Goto, ``Lyrics-to-audio alignment and phrase-level segmentation using incomplete internet-style chord annotations,'' in \emph{7th Sound and Music Computing Conference (SMC2010)}, 01 2010.

\bibitem{fujihara2011alignment}
H.~Fujihara, M.~Goto, J.~Ogata, and H.~G. Okuno, ``Lyricsynchronizer: Automatic synchronization system between musical audio signals and lyrics,'' \emph{IEEE Journal of Selected Topics in Signal Processing}, vol.~5, no.~6, pp. 1252--1261, 2011.

\bibitem{masclef2021user}
N.~L. Masclef, A.~Vaglio, and M.~Moussallam, ``User-centered evaluation of lyrics-to-audio alignment.'' in \emph{ISMIR}, 2021, pp. 420--427.

\bibitem{raffel2014mireval}
C.~Raffel, B.~McFee, E.~J. Humphrey, J.~Salamon, O.~Nieto, D.~Liang, and D.~P.~W. Ellis, ``Mir\_eval: A transparent implementation of common mir metrics.'' in \emph{Proceedings of the 15th International Conference on Music Information Retrieval}, 2014, pp. 367--372.

\bibitem{ewert2009dlnco}
S.~Ewert, M.~Muller, and P.~Grosche, ``High resolution audio synchronization using chroma onset features,'' in \emph{2009 IEEE International Conference on Acoustics, Speech and Signal Processing}, 2009, pp. 1869--1872.

\bibitem{muller2006multiscale}
M.~M{\"{u}}ller, H.~Mattes, and F.~Kurth, ``An efficient multiscale approach to audio synchronization,'' in \emph{{ISMIR} 2006, 7th International Conference on Music Information Retrieval, Victoria, Canada, 8-12 October 2006, Proceedings}, 2006, pp. 192--197.

\bibitem{pratzlich2016mr-dtw}
T.~Prätzlich, J.~Driedger, and M.~Müller, ``Memory-restricted multiscale dynamic time warping,'' in \emph{2016 IEEE International Conference on Acoustics, Speech and Signal Processing (ICASSP)}, 2016, pp. 569--573.

\end{thebibliography}
	
\end{document}